Original Article

# An Analysis Scheme for Investigation of Effects of Various Parameters on Signals in Acoustic-Resolution Photoacoustic Microscopy of Mice Brain: a Simulation Study

Hossein Ghadiri[1,2], Mohammad Reza Fouladi[*1,2], Arman Rahmim[3]

1- Medical Physics and Biomedical Engineering Department, Tehran University of Medical Sciences, Tehran, Iran.

2- Research Center for Molecular and Cellular Imaging, Tehran University of Medical Sciences, Tehran, Iran.

3- Radiology and Radiological Sciences Department, Johns Hopkins University, Baltimore, USA.

## ABSTRACT

**Purpose-** Photoacoustic spectral analysis is a novel tool for studying various parameters affecting signals in Photoacoustic microscopy. Mere observation of the frequency components of photoacoustic signals does not make enough data for a desirable analysis. Thus a hybrid time-domain and frequency-domain analysis scheme has been proposed to investigate the effects of various parameters like the depth of microscopy, laser focal spot size and contrast agent concentration on Photoacoustic signals.

**Methods-** Photoacoustic wave generation and propagation in the mouse brain phantom has been simulated using k-space method and both time-domain and frequency-domain analysis of photoacoustic signals are presented for the evaluation of three parameters affecting signals; the depth of microscopy, the concentration of Indocyanine Green as an exogenous contrast agent and the size of laser focal spot illuminating Hemoglobin as an optical absorber.

**Results-** This work demonstrates that going deeper through the brain will decrease the contrast and resolution of photoacoustic microscopy images of mouse brain vasculature and this can be resolved by using an exogenous contrast agent like Indocyanine Green. Also, by the proposed analysis of photoacoustic signals, we demonstrated that using Indocyanine Green will increase the depth resolution as well as the contrast of photoacoustic signals.

**Conclusion-** Simulations and analyses conducted on different sizes of irradiated hemoglobin absorbers demonstrated that there is a need in ultra-broadband transducers for achieving a more precise analysis of photoacoustic signals, and this means that Acoustic-Resolution Photoacoustic Microscopes are less efficient for utilization in photoacoustic spectral analysis.

## 1. Introduction

Photoacoustic Imaging (PAI) is a non-ionizing hybrid medical imaging modality based on photoacoustic effect which was described by Alexander Graham Bell at 1880 for the first time [1].

Briefly, in pulsed-mode PAI, a pulsed laser beam illuminates the tissue, then the absorption of photons, reaching the tissue Region Of Interest (ROI), leads to a slight localized heating of the tissue causing thermoelastic expansion and generating pressure waves, which can be detected by ultrasound transducer(s) as Photoacoustic (PA) signals. PAI simultaneously benefits from optical contrast and ultrasound resolution and this means more depth resolution than pure optical imaging modalities and more contrast rather than ultrasound imaging techniques, though the origins of contrast are completely different in PAI rather than imaging with ultrasound echoes in B-mode imaging. PAI, like many medical imaging modalities, utilizes agents for contrast enhancement of different targeted ROIs. These contrast agents are either endogenous or exogenous. Hemoglobin, melanin, lipid, collagen, elastin and water are some endogenous agents. Also, a variety of exogenous agents, e.g. Indocyanine Green (ICG), Evans Blue (EB), IRDye800, quantum dots and copper sulfide nanoparticles, have been used in various biomedical applications [2-8].

*Corresponding Author:
Mohammad Reza Fouladi, Msc Student
Medical Physics and Biomedical Engineering Department, Tehran University of Medical Sciences, Tehran, Iran.
Tel: 09124528066
E-mail: mrfouladi@razi.tums.ac.ir

During the last decade, PAI has found many clinical applications in fields like urology, dermatology, gynecology, hematology, ophthalmology and neuroscience studies for neuroimaging and brain mapping [9-10].

The capability of this modality for both anatomical and functional imaging has made it an opportune method for small animal brain imaging. Also, it can connect micro structural studies of brain with macro scale observations. Studies on the brain of small animals like mouse and rat have proposed methods for measuring quantities like Cerebral Metabolic Rate of Oxygen ($CMRO_2$) and Saturation of Oxygen ($SO_2$) either with or without the use of exogenous agents [11-12].

Various systems have been utilized in the literature, amongst them; Photoacoustic Microscopy (PAM) and Photoacoustic Tomography (PAT) have found major applications. Focused spherical, ring-shaped transducers, rotational multi-element transducers or novel linear array transducers have been used in PAT. Sensors detect signals of various ROIs in biological tissues and then the reconstruction of acquired PA signals forms tomographic images [13].

PAM systems are confocal microscopes which are based on their instrumentation and can be divided into two sub-systems; Optical-Resolution Photoacoustic Microscopy (OR-PAM) and Acoustic-Resolution Photoacoustic Microscopy (AR-PAM). Briefly, the focus of pulsed laser beam in OR-PAM is narrow and a broadband transducer is used for the detection of broadband ultrasound waves. In contrast, AR-PAM utilizes narrower bandwidth transducers and the focal spot of laser light is broader at the focal zone of microscopy. Broader light illumination can be achieved by the utilization of mirrors for diffusing photons [14-17].

ICG is one of the most efficient contrast agents for photoacoustic brain imaging. It is a non-toxic agent with a suitable clearance which can be transmitted across the Blood-Brain Barrier (BBB) and has an optical absorption peak at a laser wavelength of 800 nm. Wavelengths in the range of 700-1064 nm are in Near Infrared (NIR) region of optical spectrum in which light penetration achieves its maximum level, that is why almost all exogenous contrast agents are fabricated with the optical absorption peaks of wavelengths in NIR region [4,6-7].

A majority of previous studies on PAI were focused on instrumentation, different imaging techniques e.g. PAM or PAT and the implementation of novel reconstruction algorithms etc. Recently, a new technique termed Photoacoustic Spectral Analysis (PASA) has been developed. The major purpose of PASA is the quantification of different characteristics of tissues and biological microstructures by analyzing PA signal spectrums [18-19].

Earlier studies like "spectral analysis of PAI data from prostate adenocarcinoma tumors in a murine model "have focused on morphological characterization of biological tissues only by implementing analysis in frequency-domain data [20-21].

In this work, a simulation study is presented, for the first time to the best of our knowledge, both time-domain and frequency-domain analysis of PA signals generated from different concentrations of ICG. Then, PA signals from ICG have been compared with signals generated by hemoglobin in mouse brain vasculature using narrowband curved array transducers. In addition, the analysis of PA signals from different imaging depths and different sizes of illuminated optical absorbers or chromophores augment our study. Despite using AR-PAM, which is similar to our simulation setup, PASA has not been reported so far. Beside considered studies, a novel technique for the analysis of effects of different absorber sizes on PA signals has been proposed using a linear regression to power the spectrum of PA signals and evaluation of their intercept with vertical axis (y intercept).

## 2. Materials and Methods

### 2.1. Theory

When a nanosecond pulsed laser illuminates the mouse brain, light propagates through the medium and experiences several interactions; mainly absorption and scattering, i.e. refraction and reflection. The required time scale of laser pulse duration for PA wave generation can be determined using stress and thermal confinements in 1 and 2 [21].

$$T_{stress} = d/v_s \qquad (1)$$

$$T_{thermal} = d^2/4\alpha \qquad (2)$$

Where d is the characteristic dimension of the tissue being heated, in units of meter, α is the thermal diffusivity of the brain tissue in units of squared meter per second, $v_s$ is speed of sound in the brain medium in units of meter per second, $T_{stress}$ is maximum required time for the generation of pressure waves without any break-down of stress confinement condition and $T_{thermal}$ is the maximum required time for the conversion of heat to PA waves without any significant thermal diffusion.

A general model of light propagation through biological turbid media, where scattering dominates, is described by Boltzmann equation which is known as Radiative Transfer Equation (RTE) in our model. Boltzmann equation in the steady state can be written as 3 [22].

$$(\hat{s}.\nabla + \mu_t(r))\varphi(r,\hat{s}) = \mu_s(r)\int_{S^{n-1}}\Theta(\hat{s},s')\varphi(r,s')d\hat{s}'$$
$$+ q(r,s') \qquad (3)$$

Where $\mu_t(r)$ is (4)

$$\mu_t(r) = \mu_s(r) + \mu_a(r) \qquad (4)$$

Here $\mu_t(r)$, $\mu_s(r)$ and $\mu_a(r)$ are optical attenuation, scattering and absorption coefficients in the units of per meter, respectively. $\varphi(r,\hat{s})$ is the number of photons per unit volume at position r in an angular direction of velocity $\hat{s}$ in the units of per cubic meter per stradian. $q(r,s')$ is the number of source photons and $\Theta(\hat{s},s')$ is the normalized phase function corresponding to the probability of scattering from direction $\hat{s}'$ to $\hat{s}$.

Absorption of photons slightly heats the medium; then thermoelastic expansion of heated molecules generates pressure waves; this is described by Morse and Uno Ingard, Diebold and colleagues as 5 [23-24].

$$\nabla^2 p(r,t) - \frac{1}{v_s^2}\frac{\partial^2}{\partial t^2}p(r,t) = -\frac{\beta}{C_P}\frac{\partial}{\partial t}H(r,t) \qquad (5)$$

Here p(r,t) is the pressure wave in units of Pascal at position r in time t, β is the isobaric volume expansion coefficient in units of per Kelvin, $C_P$ is the isobaric specific heat and H(r,t) is the thermal energy deposited by laser illumination at position r and time t. If 1 and 2 as stress and thermal confinement conditions be met, under illumination of a nanosecond pulsed laser; the temporal component of H(r,t) can be treated as a delta function and one can approximate H(r,t) as 6

$$H(r,t) \approx A(r)\,\delta(t) \qquad (6)$$

Here δ(t) is delta function and A(r) is absorbed energy density at position r that is 7

$$A(r) = \int_{S^{n-1}}\varphi(r,\hat{s})d\hat{s} \qquad (7)$$

Equation 7 can be calculated by using 3 and 4 and then initial pressure $P_0$ generated at time $t_0 = 0$ and position r, described by Wang, is 8 [25].

$$P_0 = \Gamma\, A(r) \qquad (8)$$

Where Γ is Grüneisen parameter (dimensionless) and can be calculated by all defined parameters as 9 [26].

$$\Gamma = \beta v_s^2/C_P \qquad (9)$$

Thus, for simulating the underlying phenomena, there is a need to absorbed energy density A(r) and Grüneisen parameter Γ. As we described, for the calculation of A(r) we need to solve RTE. Different solutions of RTE have been proposed in the literature e.g. diffusion approximation, Finite Element Method (FEM) and Monte Carlo (MC) simulations are some of them. Despite considered methods, we used Neumann-series approach, described by Abhinav. K. Jha and colleagues.

Thus, results shown in Figure 1 have been used for acquiring a straightforward start point for wave equation solution in the mouse brain. Also, results of Neumann-series approach for RTE solution are compared with MC technique in Figure 1 and by detail in the literature [27].

Transmitted flux F(r) is 1 - A(r)/ $\mu_a$(r) and then A(r) is 10:

$$A(r) = \mu_a(r) (1-F(r)) \tag{10}$$

For solving wave equation in this simulation study, k-space method, described by Treeby and Cox, has been used to simulate the propagation of PA waves through medium and detection of PA signals. K-space approach could be considered as a modified pseudo-spectral technique for solving wave equations. K-space method modifies the standard differencing method for time integration by introducing a periodic function, so that much larger time steps can be chosen without introducing inaccuracy and instability. Therefore, it results in saving the computation time and memory significantly compared to other numerical methods [28-29].

### 2.2. Simulation

An acoustically heterogeneous mathematical mouse brain phantom has been developed in this study. A sphere with radius of 5 mm as mouse head, the skull with uniform thickness of 0.5 mm and the brain tissue with radius of 4.5 mm have been simulated in a FDTD grid-based medium. Figure 2 shows numerical phantom with spatial resolution of 230 μm and focused concave array sensor used. Materials used in our simulation study were water, mouse skull, brain, hemoglobin and ICG, which their optical and acoustical properties, for numerical study, are listed in Table 1. Also, we should mention that water has been used for the acoustical coupling of sensor and mouse brain.

**Table 1.** Acoustical and optical properties of simulated materials.

|  | Attenuation Coefficient dB/MHz.cm | Sound Speed m/s | Density g/cm$^3$ | Extinction Coefficient ε (cm$^{-1}$M$^{-1}$) |
|---|---|---|---|---|
| **Water** | 0.002 | 1480 | 1 | - |
| **Skull** | 20 | 4180 | 190 | - |
| **Brain** | 0.8 | 1550 | 103 | - |
| **Hemoglobin** | - | - | - | 816 |
| **ICG** | - | - | - | 154550 |

At the first step, we have acquired PA signals from spherical optical absorbers containing hemoglobin with radiuses of 234 μm, 468 μm, 702 μm and 936 μm while the illumination of constant laser fluxes with different focal spot sizes which illuminates a considered area of hemoglobin. At the second step, we have acquired PA signals from ICG concentrations of 0.5 mg/L, 0.5 g/L, 5 g/L and 50 g/L. At the third step, we have acquired PA signals from hemoglobin at depths of 0.6 mm, 0.8 mm and 1 mm under internal surface of the mouse skull i.e. 1.1 mm, 1.3 mm and 1.5 mm under the external surface of mouse head.

A pulsed laser with pulse duration of 5 ns at wavelength of 800 nm with an initial energy flux of 31.7 mJ/cm$^2$ illuminated the mouse brain from a point with normal distance of 0.25 mm to the external surface of mouse head.

Any exposure to biological tissues should follow radiation safety considerations and in this study, Maximum Permissible Exposure (MPE) on skin should follow ANSI standard which for pulsed lasers with wavelengths in the range of 700-1050 nm is 11

$$MPE = 20 \times 10^{2(\lambda-700)/1000} \tag{11}$$

Here, λ is wavelength in units of nanometer and MPE is in units of mili-Joule per square centimeter [30].

Thus, at wavelength of 800 nm in this simulation MPE could be 31.7 mJ/cm$^2$ of which value as initial laser energy flux has been used in this work.

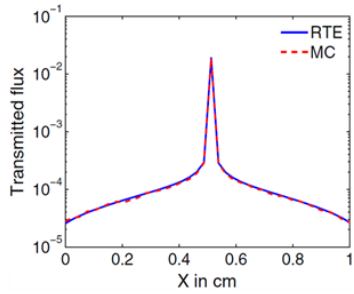

**Figure 1.** Neumann-series approach for RTE solution and Monte Carlo (MC) simulation method results in a biological turbid media [15].

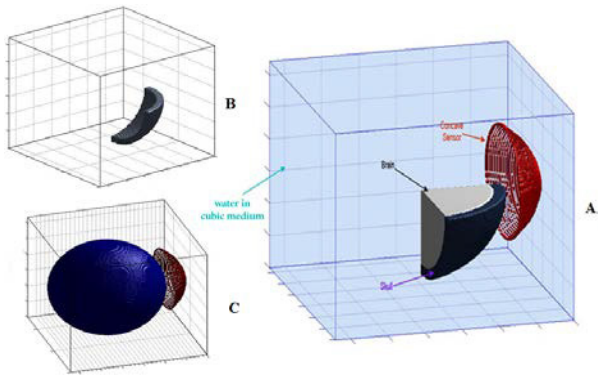

**Figure 2.** Representation of mathematical mouse brain phantom including soft tissue and skull and focused sensor used in simulations. A: a section of mouse brain in simulation setup B: skull section C: schematic of complete simulation

Meanwhile, Grüneisen parameter has been assumed with a value of 0.8 as an average value for mouse brain as a fat tissue according to the literature [31].

A concave focused array sensor, focusing at different depths of $P_0$ generation, as ultrasound transducer has been used. The concave sensor consists of 97 point sensors and pressure data has been averaged over all sensors in all measurements. The maximum frequency of sensor is 3 MHz and geometrically it is a section of a sphere with a radius of 1.4 mm. The pulsed laser and concave focused sensor have been installed in a confocal manner and simulated sensor has a uniform bandwidth of 3 MHz. As discussed, this simulation setup is similar to an AR-PAM system setup.

## 2.3. Analysis of Photoacoustic Signals

"Analysis" term has not been considered dedicated to frequency-domain analysis and power spectrum measurements. In addition, we discuss the time-domain PA signals and their amplitudes in considered situations for the evaluation of various parameters affecting PA signals.

### 2.3.1. Time-Domain Analysis

All peak-to-peak pressures (PPP) of PA signals at the sensor position have been measured and compared. It has been assumed that transducer would detect PPP and converts it to voltage with a high sensitivity, and then it can be amplified using an amplifier. Simply, PPP at the sensor position $r_0$ can be calculated as following (12)

$$PPP(r_0) = P_{max}(r_0) - P_{min}(r_0) \qquad (12)$$

Where, $P_{max}(r_0)$ and $P_{min}(r_0)$ are maximum pressure and minimum pressure detected at the sensor position $r_0$, respectively. One should consider that a positive pressure means an expansion and a negative pressure means a contraction in bipolar PA signals of a single spherical $P_0$ source.

### 2.3.2. Frequency-Domain Analysis

Discrete-time Fourier transformation has been done using Fast Fourier Transform (FFT) algorithm which is described by details in the literature [32]. Also, Power Spectral Density (PSD) of acquired signals in considered situations are estimated.

Theoretically, scaling the magnitude squared of the discrete-time Fourier transform of signals is an estimate of PSD; if we assume a signal at sensor position $r_0$ and in the n-th time step of simulation with a pulse duration of $\tau = 5$ ns, then there is (13).

$$PSD(f) = \frac{1}{(t_{end})F_S} \left| \sum_{n=0}^{t_{end}-1} p(r_0, n\tau) e^{-2\pi i fn/F_S} \right|^2 \qquad (13)$$

Where $F_S$ is the sampling frequency of discrete-time Fourier transform and $t_{end}$ is the last time step of simulations which would change during different circumstances and a typical value can be 10 microseconds. In our study, we have used a method for the estimation of PSD which applied the fast Fourier transform algorithm and involves sectioning the record, taking modified periodograms of these sections, and averaging these modified periodograms [33].

### 2.4. Analysis of Effects of Illuminated Hemoglobin Size on PA Signals

In the analysis of illuminated absorber size effects on PA signals, we demonstrated that linear regression of two first sequential maximums and their intercept with power spectrum vertical axis can help quantify the size effects on acquired PA signal power spectrums. After the calculation of y intercepts, a plot of power spectrum y intercept as a function of absorber size has been plotted and a function of the form 14 has been fitted to the data for evaluating the effects of size on PA signals.

$$\text{psyi} = c_1(s)^\gamma + c_2 \quad (14)$$

Here, psyi is y intercept of linear regression to two first maximums of power spectrums, $c_1$, $c_2$ and $\gamma$ are parameters which have been calculated from simulation results and s is the size of illuminated absorbers by different laser focal spot sizes.

### 2.5. Analysis of Effects of ICG Concentration on PA Signals

Increasing the concentration of any agent, from a physical point of view, means an increase in initial pressure generated while heating the tissue under pulsed laser illumination. Thus, ICG concentration increase does not affect the frequency components of PA signals but utilizing an exogenous contrast agent like ICG can affect the frequency components of PA signals relative to an endogenous contrast agent like hemoglobin.

For an analysis of effects of ICG injection on generated PA signals, power spectrums of equal concentrations of ICG and hemoglobin have been achieved. Then, the average power of spectrum in array transducer frequency range which is known as band power is a measure for evaluation of effects of ICG existence in brain vasculature.

Theoretically, Band Power (BP) can be calculated by 15

$$BP = (1/Fs) \int_{fr} PSD(f) df \quad (15)$$

Here, fr is frequency range (0 – 3 MHz). BP can be used as a measure for power spectrum investigations [34].

### 2.6. Analysis of Effects of Imaging Depth on PA Signals

For the analysis of depth effects on PA signals, higher frequency components of power spectrum have been studied and numerical integration of power spectrums in three different depths of imaging in the frequency range of 1.5-3 MHz have been calculated using rectangle approximation to power spectrum curves.

## 3. Results

### 3.1. Analysis of PA Signals From Different Illumination Sizes of Hemoglobin

Figures 3-5 are analyses results of PA signals from Hemoglobin with the illumination focal spot dimensions of 234 μm, 468 μm, 702 μm and 936 μm.

Figure 3 shows the relationship between PPP and size. In Figure 3 PPPs are $1.3 \times 10^{-16}$ Pa, $2.6 \times 10^{-16}$ Pa, $5.7 \times 10^{-16}$ Pa, $12.5 \times 10^{-16}$ Pa for illumination radiuses of 234 μm, 468 μm, 702 μm and 936 μm, respectively. Figure 4 shows the power spectrum density estimates of considered sizes. In Figure 4, there is a frequency component of 1.75 MHz in the spectrum of 234 μm which cannot be resolved in spectrums of other sizes. Figure 5 shows both simulation data and fitted line for power spectrum y intercept as a function of hemoglobin size. As discussed, fitted line follows Equation 14. According to the fitted line using Equation 14, values of $c_1$, $c_2$ and $\gamma$ are $1.0932 \times 10^{-45}$, $-1.2470 \times 10^{-35}$ and 4.329, respectively.

### 3.2. Analysis of PA Signals From Different Concentrations of ICG

Figures 6-8 are the results of the analysis of PA signals from ICG concentrations of 0.5 mg/L, 0.5 g/L, 5 g/L and 50 g/L. Figure 6 shows a linear relationship between ICG concentration and PPP of PA signals using time-domain analysis of PA signals. Figure 7 shows time-domain PA signals of the same concentrations of hemoglobin and ICG. In Figure 7, maximum signal amplitude of hemoglobin is about $5.5 \times 10^{-16}$ Pa which is about $2 \times 10^{-13}$ Pa for ICG. Figure 8 shows the power

spectrums of ICG and hemoglobin with equal concentrations. In Figure 8, there is a frequency component about 1.5 MHz in ICG spectrum which cannot be resolved in the spectrum of hemoglobin. Also, estimated band power of ICG is $1.7\times10^{-30}$ which it's $2.1\times10^{-35}$ for hemoglobin.

### 3.3. Analysis of PA Signals From Different Depths of Microscopy

Figures 9-12 are the results of the analysis of PA signals from imaging depths of 1.5 mm, 1.3 mm and 1.1 mm. Figure 9 shows time-domain PA signals from different depths of microscopy. In Figure 9, the sensing time of maximum PA wave's amplitudes are 1.34 µs, 1.525 µs and 1.68 µs for depths of 1.1 mm, 1.3 mm and 1.5 mm, respectively. Figure 10 shows the relationship between PPP and the depth of imaging in which a decrease in PPP with increasing depth is obvious. Figure 11 shows the power spectrums of PA signals generated from different depths and Figure 12 shows the result of power spectrums integration of 1.1 mm, 1.3 mm and 1.5 mm depths over frequency ranges of 1.5-3 MHz as a function of imaging depth.

## 4. Discussion

### 4.1. Effects of Illuminated Hemoglobin Size on PA Signals

Hemoglobin as an optical absorber with illumination radiuses of 234 µm, 468 µm, 702 µm and 936 µm have been studied in the depth of 1.5 mm and concentration of 150 grams per liter in plasma of blood in mouse brain vasculature.

Figure 3 shows that when the size of spherical hemoglobin containers increases, an increase in PPP can be seen. This means with increasing size of microscopy field, stronger PA signals are achieved, though in reality; when the size increases other PA sources like subcutaneous melanoma will appear which can affect signals from other sources. That is why all PAM experiments for brain vasculature imaging usually utilize small animals with removed scalps.

Figure 4 shows when the size of microscopy field decreases, the center frequency of detected PA signals will increase and this means that the imaging of smaller objects can be done using broadband ultrasound transducers much better, although we have gone through this problem using a sensor with a uniform bandwidth of 3 MHz. Another peak about the frequency of 1.75 MHz in the graph of absorber with a radius of 234 µm in power spectrum of Figure 4 shows that if the maximum frequency of detection system increases, then more peaks about higher frequencies is seen and this shows the nature of broadband PA signals; this phenomenon was not seen in larger sizes because of wave interferences in macroscopic scale.

Figure 5 shows when the size of illumination increases, power spectrum y intercept will increase. Based on a fitted line obeying Equation 14, the proposed method, using y intercept for the evaluation of illumination sizes effects, can make the formulation 16 for the prediction of y intercept in linear regression of two first maximums in power spectrum. This prediction technique can be a methodology for a quantitative photoacoustic microscopy and evaluation of different illuminated absorber sizes.

$$\text{psyi} = (1.0932\times10^{-45}(s)^{4.329}) - (1.2470\times10^{-35}) \qquad (16)$$

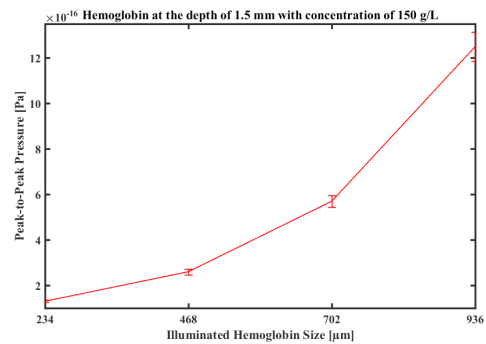

**Figure 3.** PPP as a function of hemoglobin size.

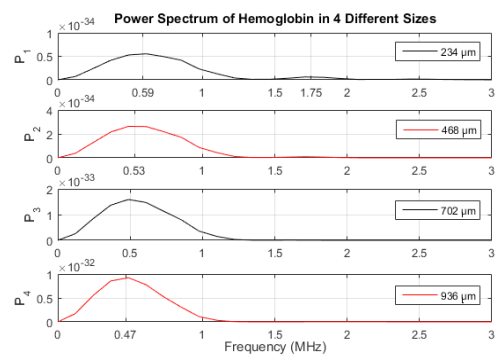

**Figure 4.** Power spectrums of hemoglobin with sizes of 234 µm, 468 µm, 702 µm and 936 µm.; existence of higher frequencies in P1 (234 µm) is obvious.

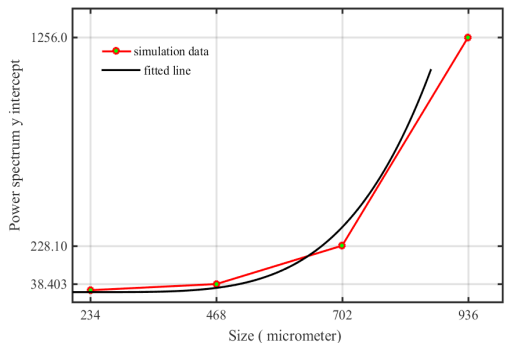

**Figure 5.** Power spectrum y intercept as a function of hemoglobin size.

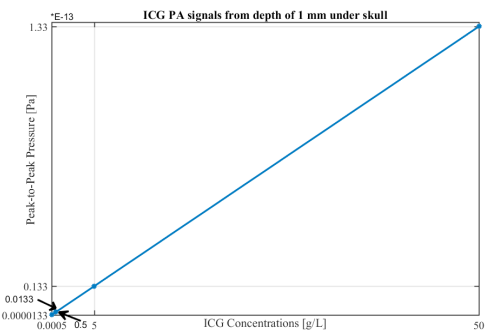

**Figure 6.** PPP as a function of ICG concentration.

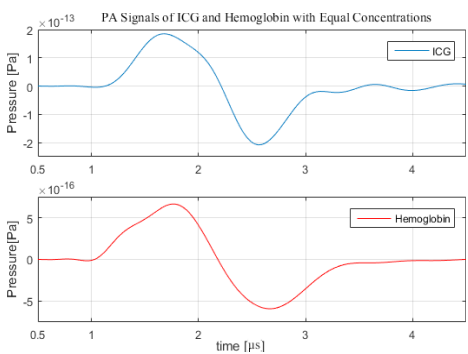

**Figure 7.** Time-domain PA signals from similar concentrations of hemoglobin (red) and ICG (blue).

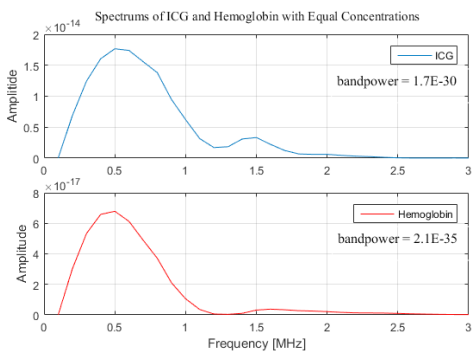

**Figure 8.** Power spectrums of ICG (blue) and hemoglobin (red) with equal concentrations.

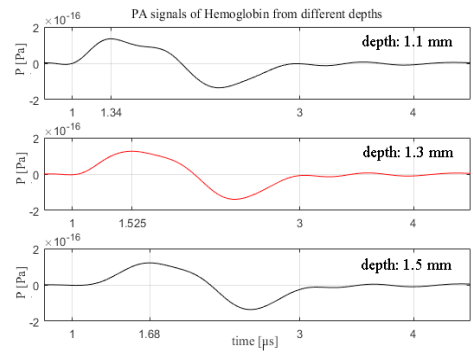

**Figure 9.** Time-domain PA signals of hemoglobin from depths of 1.1 mm, 1.3 mm and 1.5 mm.

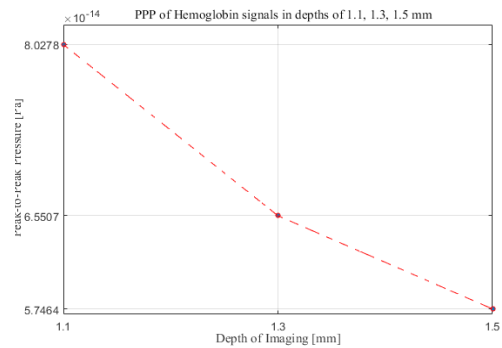

**Figure 10.** PPP as a function of PAM depth.

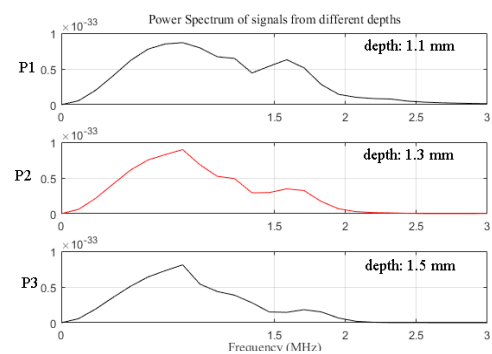

**Figure 11.** Power spectrums of PA signals from depths of 1.1mm, 1.3mm and 1.5mm.

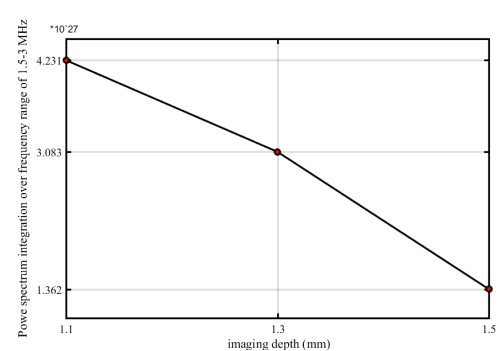

**Figure 12.** Power spectrum integration of hemoglobin over frequency range of 1.5-3 MHz as a function of imaging depth.

### 4.2. Effects of ICG Concentration on PA Signals

ICG concentration affects optical absorption coefficient, then $P_0$ amplitude will increase and with higher concentrations of ICG in brain vasculature one can detect stronger waves at the position of sensor. Figure 6 shows an increase in PPP with increasing concentration of ICG which can be measured with A(r), $P_0$, depth and acoustic attenuation of mouse brain and this means, when ICG concentration increases, SNR will be increased effectively and higher concentration ICGs will become contrast origin in PAM at wavelength of 800 nm. Although, one should consider that The $LD_{50}$ after IV administration ranges between 50 and 80 mg/Kg for ICG in mice [35].

Figures 7-8 show differences between PA signals of the same concentrations of Hemoglobin and ICG both in time-domain and frequency-domain, respectively. Figure 7 shows that PA signal amplitude of ICG is 1000 times stronger than PA signal amplitude with equal concentrations and this shows the ability of ICG to make higher SNRs relative to the most important endogenous contrast agent (Hemoglobin).

Figure 8 shows another peak about the frequency of 1.5 MHz and this reveals the capability of ICG in making greater resolutions in PAM systems, in addition to its contrast enhancement. Also, the band power of ICG signals is much higher than the band power of hemoglobin signals, which augments our founding for SNR enhancement and resolution improvement of ICG injection.

### 4.3. Effects of Microscopy Depth on PA Signals

Figure 9 indicates when the depth of microscopy increases, the signal acquisition time will increase and reverberations of PA signals will interfere and this means that going through deeper tissues in mouse brain would corrupt the real-time capability of PAM, especially in functional studies and more processings are needed for signal detection.

Figure 10 indicates that the imaging of deeper tissues will decrease contrast density. This issue specifies when thermal or random noises are added to the situation; PA signal cannot be resolved and SNR will be dramatically decreased. In addition to hemoglobin contrast reduction, Figure 11 shows going thorough deeper vasculature in mouse brain, the depth resolution will decrease. In 1.1 mm under surface of head, frequency components of 2.5 MHz can be detected while they cannot be resolved in 1.5 mm imaging depth.

Figure 12 shows when imaging depth increases, the integral of power spectrum over frequency range of 1.5-3 MHz will decrease and this concept makes a better understanding of depth resolution reduction when going deeper through brain. One of the most challenges of photoacoustic imaging is its penetration depth which should be studied more for going deeper through tissues without losing much SNR and resolution.

### 5. Conclusion

We have presented both time-domain and frequency domain analysis of PA signals from various imaging depths, different sizes of illumination and various ICG concentrations in mouse brain vasculature.

We have demonstrated that going deeper through the mouse brain will decrease the contrast and resolution of our simulated PAM system, something which can be solved by using exogenous contrast agents with a greater optical absorption coefficient such as ICG. Comparison of PA signal amplitudes of ICG and Hemoglobin shows a great demand for using various contrast agents in PAM, because, as we discussed, this will increase both depth resolution and contrast of this microscopic modality.

A novel method for analysis of the effects of absorber sizes under illumination on power spectrum has been proposed. Under some circumstances, Eequation 16 can provide useful data for underlying situations.

Finally, simulations and analysis undertaken on different sizes of illuminated absorbers in PAM demonstrated that there is a need to ultra-broadband transducers for reaching more precise analysis of PA signals, and this means AR-PAM systems are less efficient for being used in photoacoustic spectral analysis.